# Modeling the Dielectric Relaxation in Semicrystalline Polymers – Understanding the Role of the Interphase between the Amorphous and Crystalline Domains

Valeriy V. Ginzburg[*]

Department of Chemical Engineering and Materials Science, Michigan State University,

East Lansing, MI 48824

Corresponding author, email ginzbur7@msu.edu

# Abstract

Semicrystalline polymers (SCP) represent important class of materials used in many applications from packaging to transportation to electronics to pharmaceuticals. Understanding the structure and dynamics of the crystalline and amorphous fractions within SCP is an important challenge for both experimentalists and theoreticians. Recently, Cheng et al. (S. Cheng et al., J. Chem. Phys. 160, 114904 (2024)) utilized Broadband Dielectric Spectroscopy (BDS) to explore the relaxation time profiles within the amorphous region of a semi-crystalline poly(L-lactic acid) (PLLA) with ordered alternative stacking of crystalline/amorphous phases. Here, we build on those initial results and model the relaxation time distribution using the combination of the Doolittle free volume theory (FVT) and the compressible Self-Consistent Field Theory (cSCFT). We show that the new modeling framework successfully captures the temperature and position dependence of the local relaxation time of the semicrystalline PLLA above the glass transition temperature of the amorphous region. Finally, we derive a new empirical equation for the relaxation time profiles in amorphous regions of SCPs.

## 1. Introduction

Broadband dielectric spectroscopy (BDS) is an important technique that probes the material dynamics and relaxation times over a wide range of temperatures and time scales.[1–5] For example, the dependence of the dielectric $\alpha$-relaxation time on the temperature exhibits a strong upward nonlinearity from a traditional Arrhenius behavior in the vicinity of the glass transition; this behavior can be modeled using empirical equations like Vogel-Fulcher-Tammann-Hesse (VFTH),[6–8] Williams-Landel-Ferry (WLF),[9] Avramov-Milchev,[10] Mauro and co-workers (MYEGA),[11] Ginzburg (TS2),[12–15] Elmatad *et al.*[16] and others. On a fundamental level, glass transition is variously described based on "free volume",[17,18] "configurational entropy,"[19–21] Mode-Coupling Theory,[22–28] the Elastically Collective Nonlinear Langevin Equation (ECNLE) model,[29–35] and other approaches. At temperatures below the glass transition temperature, $T_g$, but above the so-called Kauzmann temperature, $T_K$, the free volume or the configurational entropy nearly vanish and the relaxation time increases many orders of magnitude.

The problem becomes even more challenging for heterogeneous systems like thin films (free-standing and supported), polymer-particle nanocomposites, and semi-crystalline polymers.[36–52] The mobility of polymers at interfaces can be dramatically enhanced (near free surfaces) or inhibited (near hard surfaces), leading to the overall relaxation time reduction (for free-standing films) or increase (for nanocomposites and supported films). Simulations and theory[33,34,53–61] have been widely used to understand these phenomena, although many questions still remain.

Recently, Cheng and co-workers[62] used broadband dielectric spectroscopy to probe the relaxation dynamics of polymer under hard nanoconfinement of around 10 nm. Different from previous measurements on support or capped thin films, the authors took advantage of melt pre-deformation followed by a controlled thermal annealing of a semicrystalline poly(L-lactic acid) (PLLA), which produces an ordered lamellar structure of alternating stacking crystalline/amorphous phases. Since the amorphous polymer phase is sandwiched by two neighboring crystalline phases, the lamellar structure thus mimics a stacking of nanoconfined capped polymer film. Due to the unique sample preparation procedure, the amorphous polymer phase is mostly composed of trapped entanglement or tie chains with chain ends anchoring on the neighboring crystalline phase. Given that the polymers in the crystalline phase are significantly less mobile than their amorphous counterparts, the dielectric measurements can selectively capture the dynamics contribution of the amorphous phase within the frequency window of the measurements. Interestingly, the dielectric properties of the confined amorphous polymers of the semi-crystalline polymers exhibit strong low-frequency broadening that increases with cooling, while the stretching parameter of the high-frequency-side remain unchanged. These results demonstrate a clear interface effect on dynamics of amorphous polymer phase. Furthermore, the authors developed a new procedure, "continuous" Havriliak-Negami (cHN) analysis, to deconvolute the dielectric spectra and estimate the dependence of the "local" relaxation time on the distance from the crystalline/amorphous interface. A power-law decay of the relaxation time gradient has been obtained at the 1-2 nm vicinity of the crystalline/amorphous interface before the spatial gradient of the relaxation time changes to a single or double exponential spatial distribution. The authors attribute the deviation from theory mostly from the unique interfacial chain packing, i.e. high-density surface chain-anchoring due to the large populations of the trapped entanglement or tie chains. These results thus emphasize a role of the detailed chain packing at the interface on the interface dynamics gradient. Despite the significant differences in the chain conformation of the semi-crystalline PLLA from the thin film prepared through spin coating, these profiles are important data points for theoretical models aiming to describe the dynamics of heterogeneous glass-formers.

In terms of theory and modeling, Karakus and co-workers[63] recently formulated a simple free-volume theory describing the dependence of the "average" relaxation time in a glycerol-silica nanocomposite on the temperature and the silica nanoparticle loading. Starting from the equilibrium density profile of glycerol near the silica surface, computed using atomistic Molecular Dynamics, they coarse-grained it to obtain a smooth monotonic function and interpreted it in terms of free volume. The "free volume profile" was then converted into the "relaxation time profile". Using logarithmic averaging procedure, they subsequently obtained the overall relaxation time as a function of temperature and "average interparticle distance" (which was then translated into the particle volume fraction). The calculations agreed well with experimental data of Cheng et al.[64]

The goal of this study is to extend the earlier model to describe the relaxation time profiles for semi-crystalline polymers, treating the crystalline phase similar to fillers in nanocomposites. Thus, we compute the coarse-grained density profiles of the amorphous polymers within the nanoscale domains between the adjacent crystallites. This calculation is performed using compressible Self-Consistent Field Theory (cSCFT), so the resulting density profiles reflect the polymeric nature and the unique surface feature of the semi-crystalline PLLA (for example, the fact that the chains could either bridge the two adjacent crystallites or form loops between two points on the same crystallite surface). We then compute the relaxation time profiles and compare them with the ones determined from the dielectric spectroscopy data using the cHN approach.

## 2. Materials and Methods

### 2.1. Semi-crystalline PLLA with stacking crystalline/amorphous phases.

Recent studies demonstrate the access of an interesting morphology of semi-crystalline polymers with alternating crystalline and amorphous phases, where the sizes of both the crystalline phase and the amorphous phase have narrow distribution.[62,65,66] This interesting structure is prepared through polymer pre-deformation close to their glass transition followed by a controlled thermal annealing for cold recrystallization. Although the molecular mechanism for the structure formation still requires further investigation, it is believed the crystallization of the pre-deformed polymer between neighboring topological entanglement points locks the entanglements and produces a lamellar-like structures with alternating crystalline and amorphous phases.[62,65,66] The amorphous phase should be mainly composed of trapped entanglements or tie chains. For PLLA, the size of the amorphous phase was measured to be 10.6 nm and the crystalline phase 6.7 nm with narrow sizes distribution,[62] enabling a platform to access of polymer dynamics under ultrafine nanoconfinement down to around 10 nm. The amorphous domains, thus, are similar to nanoconfined thin films, and the dielectric measurements can selectively probe their dynamics due to the nearly frozen mobility of the crystalline phase. The details of the sample preparation and characterizations can be found in the article by Cheng et al.[62] Here, we build on their analysis of dielectric measurements of the semi-crystalline PLLA with lamellae thickness of 6.7 nm (polydispersity index of 1.03) and amorphous phase of 10.6 nm (polydispersity index of 1.01).

### 2.2. The relaxation time distribution and the spatial gradient of segmental dynamics of the amorphous phase in semi-crystalline PLLA from broadband dielectric spectroscopy measurements

Here, we briefly summarize the analysis of Cheng et al.[62] that translates the dielectric spectrum into the relaxation time profile within the amorphous domains. The total dielectric properties of the semi-crystalline PLLA with alternating crystalline/amorphous phases, $\varepsilon_t^*(\omega)$, are composed of the contributions of the crystalline phase, the amorphous phase, the dielectric constant at infinitely high frequency, $\varepsilon_\infty$, and the dc-conductivity, $\sigma_{dc}$,

$$\varepsilon_t^*(\omega) = (1-\varphi_c)\varepsilon_a^*(\omega) + \varphi_c \varepsilon_c^*(\omega) + \varepsilon_\infty + \frac{\sigma_{dc}}{i\omega\varepsilon_0} \qquad (1)$$

where $\varepsilon_a^*(\omega)$ and $\varepsilon_c^*(\omega)$ are the complex dielectric permittivity of amorphous PLLA and crystalline PLLA, respectively, $\varphi_c$ is the volume fraction of the crystalline phase, $\varepsilon_0$ is the vacuum permittivity, $\omega$ is the angular frequency, and $i$ is the imaginary unit. Since no active molecular relaxation processes of the crystalline phase locate in the temperature and frequency window of the measurements, $\varepsilon_c^*(\omega)$ can be taken as a constant. Thus, one can define $C \equiv \varphi_c \varepsilon_c^*(\omega) + \varepsilon_\infty$ as a constant. The dielectric contribution of the amorphous phase can be written as:

$$\varepsilon_a^*(\omega) = \frac{1}{d_a} \int_0^{d_a} \varepsilon^*(z,\omega) dz = \int_{-\infty}^{+\infty} \varepsilon^*(z,\omega) g\left(\ln[\tau_z]\right) d\ln[\tau_z]$$

$$= \int_{-\infty}^{+\infty} \frac{\Delta \varepsilon_z}{\left[1+\left(i\omega \tau_{HN,z}\right)^{\beta_z}\right]^{\gamma_z}} g\left(\ln[\tau_z]\right) d\ln[\tau_z]$$

(2)

where $d_a$ is the size of the amorphous phase, $\varepsilon^*(z;\omega)$ is the complex dielectric function of the amorphous layer at distance $z$ away from one crystalline/amorphous interface, $\tau_z$ is the relaxation time of the amorphous phase at $z$ distance from one crystalline/amorphous interface. In the second equal sign, we use $dz/d_a = g(\ln(\tau_z))d(\ln(\tau_z))$ representing the volume fraction of amorphous layer at distance $z$ away from one crystalline/amorphous interfaces and $g(\ln[\tau_z])$ is the relaxation time distribution density function of the amorphous polymer with relaxation time $\tau_z$. The normalization criteria holds as $\int_{-\infty}^{+\infty} g(\ln[\tau_z])d(\ln[\tau_z]) = 1.0$. We adopt $\varepsilon^*(z;\omega) = \frac{\Delta \varepsilon_z}{(1+(i\omega\tau_{HN,z})^{\beta_z})^{\gamma_z}}$, the Havariliak-Negami (HN) function describing dielectric relaxation processes, where $\Delta\varepsilon_z$ is the dielectric relaxation amplitude, $\tau_{HN,z}$ is the HN time, $\beta_z$ and $\gamma_z$ are the HN shape parameters. The $\tau_{HN,z}$ and $\tau_z$ are related through $\tau_{HN,z} = \tau_z \left[tan\left(\frac{\pi}{2(\gamma_z+1)}\right)\right]^{1/\beta_z}$.

In the analysis, we predetermine $\Delta\varepsilon_z = \Delta\varepsilon$, $\beta_z = \beta$ and $\gamma_z = \gamma$ from the dielectric function at high temperatures where the interface effect is strongly suppressed along a dielectric envelope. With these input, one can deconvolute $\varepsilon_t^*(\omega)$ through combining eq 1 and eq 2 through generalized regulation method[67] to compute $\Delta\varepsilon * g(ln\tau_z)$, from which the contribution of dielectric response at $z$ distance, $\tau_z$, to the total dielectric spectra can be obtained. As a result, $\tau_z$, the spatial distribution of the segmental relaxation time, can be obtained. The details of how to obtain $\tau_z \sim z$ from $\Delta\varepsilon * g(ln\tau_z)$ can be found in Cheng et al.[62] Below we focus on the recent analysis result of $\tau_z \sim z$ at different temperatures of a semi-crystalline PLLA with the thickness of amorphous phase of 10.6 nm.

### 2.3. Overview of the Combined Model

To model the relaxation time dependence on both temperature and position in the amorphous PLLA domain, we develop a combined approach shown in Figure 1. The complex dielectric permittivity of PLLA, $\varepsilon^* = \varepsilon' - i\varepsilon''$, is measured at multiple temperatures and frequencies, as discussed above. The resulting spectra are then analyzed using the "continuous" Havriliak-Negami (HN) model developed by Cheng et al.[62] and thus converted into the "relaxation time profiles" $\tau_\alpha(z,T)$, where $z$ is the distance from the nearest crystal-amorphous interface, and $T$ is the temperature. We then apply the "naïve" Doolittle[17]-style free-volume theory (FVT) to compute the effective fractional free volume, $\phi_v$, as a function of temperature and position. Finally, we describe the free-volume profile using compressible Self-Consistent Field Theory.[68–76] Below, we discuss the individual models in more detail.

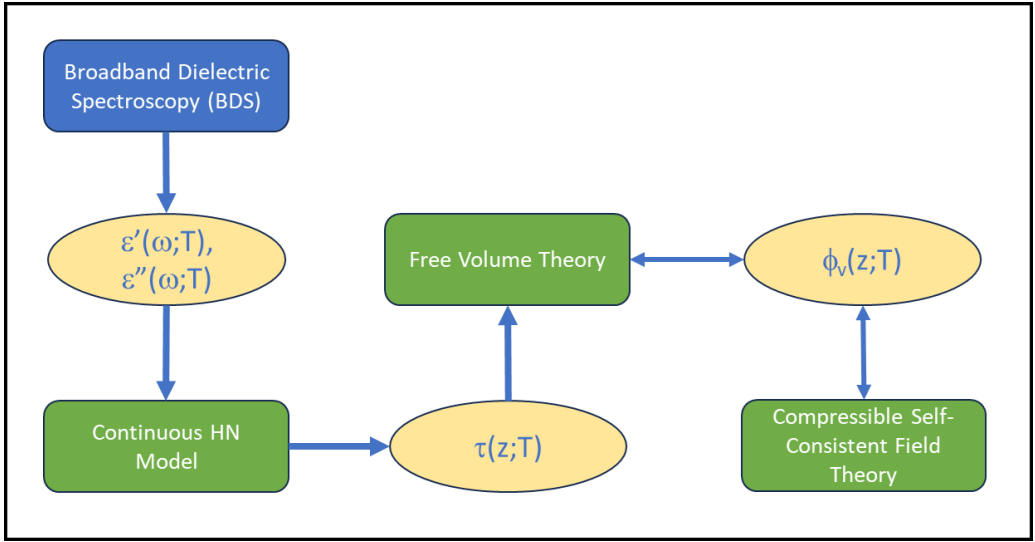

*Figure 1. Flowchart for the combined model developed to study the structure and dynamics of SCP. For more details, see text.*

### 2.4. Free Volume Theory

Free volume theory (FVT) is a well-established theory describing the relaxation dynamics of supercooled liquids (including polymers) near their glass transition temperatures.[17,18,77] A "naïve" FVT can be described as follows. The α-relaxation time, $\tau_\alpha$, as a function of temperature, $T$, can be described by a Vogel-Fulcher-Tammann-Hesse (VFTH)[6–8] equation,

$$\ln \frac{\tau_\alpha(T)}{\tau_\infty} = \frac{B}{T - T_0} \qquad (3)$$

Here, $B$, $T_0$, and $\tau_\infty$ are material-dependent parameters. Equation 3 is generally applicable to temperatures close to the glass transition temperature, $T_g$, and fails at higher temperatures, $T > 1.2 T_g$, where a more conventional Arrhenius expression applies.[78] Typically, $T_0 \approx T_g - 50\ K$. Obviously, equation 3 predicts the divergence of the relaxation time (or viscosity) as T → $T_0$ from above. This divergence is now actively disputed, and multiple non-divergent models have been proposed as alternatives.[10–12,79] However, our current study is restricted to temperatures above $T_g$, so the use of the VFTH equation is justified.

Within the "naïve" free volume theory (FVT), the α-relaxation time is given by,

$$\ln \frac{\tau_\alpha(T)}{\tau_\infty} = \frac{A}{\phi_v} \qquad (4)$$

Here, $A$ is a material-dependent parameter, and $\phi_v$ is the fractional free volume. By comparing equations 1 and 2, it follows that,

$$\phi_v = \frac{A}{B}(T - T_0) \qquad (5)$$

The "naïve" FVT assumes that the relative increase in the fractional free volume is identical to the relative increase in the observable specific volume, thus $A/B = \alpha$, where $\alpha$ is the equilibrium volumetric coefficient of thermal expansion. While this assumption obviously breaks down below $T_g$, it is reasonable for $T > T_g$. (For ways to go beyond "naïve" FVT, see Lipson and White[18,56–58,61,80–82] or Ginzburg et al.[12,14,15,83]; the derivation of the "naïve" FVT as a limiting case of a more elaborate, two-state FVT is provided by Karakus et al.[63])

Let us now consider an inhomogeneous system in which the free volume (i.e., the density) varies with position. In that case, eq 3 no longer holds; however, we assume that eq 4 is still applicable, so,

$$\ln \left[\frac{\tau_\alpha(z;T)}{\tau_\infty}\right] = \frac{A}{\phi_v(z;T)} \qquad (6a)$$

$$\phi_v(z;T) = \frac{A}{\ln \left[\frac{\tau_\alpha(z;T)}{\tau_\infty}\right]} \qquad (6b)$$

Thus, given a density profile near the surface (determined from molecular simulations or mesoscale theories), one can estimate the relaxation time distribution (eq 6a); alternatively, given the relaxation time distribution, one can compute the density profile (eq 6b).

### 2.5. Compressible Self-Consistent Field Theory

We assume that the free volume profile is the equilibrium one, as the system we study is above the glass transition temperature. To estimate the equilibrium profile, we use compressible Self-Consistent Field Theory, as originally developed by Hong and Noolandi[68] on the basis of the lattice model of Sanchez and Lacombe.[75,76,84] For simplicity, and to avoid dealing explicitly with the interfaces, we model the polymer as a triblock copolymer CAC, where C corresponds to the crystalline portion, and A to the amorphous one. The total chain length or degree of polymerization is denoted $N$. Because of the triblock architecture, no "dangling ends" are expected to be found in the amorphous domain – all amorphous chains are either bridges ("tie chains"[85]) or loops.[86]

One crucial question is – how to properly represent the free volume voids. Conventionally, the voids were modeled as a "solvent" – each hole element occupied one lattice site.[68–71] This would imply that the holes are uncorrelated – the local free volume fraction depends only on the type of the monomers surrounding it. Recently, however, Douglas and co-workers suggested that the high-mobility regions form highly correlated "strings";[87–90] in our free-volume language, it would imply that a better representation of the free volume holes would by as "homopolymers". In that case, the free volume "component" would be modeled as a string of connected lattice elements with the total length $M$. Our system thus becomes similar to a copolymer/homopolymer blend where the additive "homopolymer" (free volume elements) have a strong preference to one block (the amorphous one). Such selective copolymer/homopolymer mixtures have been extensively modeled in the past, including with SCFT.[91–94] The free energy of the mixture is given by,

$$\frac{Fv_0}{Vk_BT} = \frac{\varphi_{PLLA}}{N} \ln\left[\frac{V\varphi_{PLLA}}{Q_{PLLA}}\right] + \frac{\varphi_V}{M} \ln\left[\frac{V\varphi_V}{Q_V}\right]$$

$$+ \frac{1}{V}\int_V d\mathbf{r}\left[\chi_{AC} N(\Phi_A(\mathbf{r}) - \phi_A)(\Phi_C(\mathbf{r}) - \phi_C)\right]$$

$$+ \frac{1}{V}\int_V d\mathbf{r}\left[\chi_{AV} N(\Phi_A(\mathbf{r}) - \phi_A)(\Phi_V(\mathbf{r}) - \phi_V)\right]$$

$$+ \frac{1}{V}\int_V d\mathbf{r}\left[\chi_{CV} N(\Phi_C(\mathbf{r}) - \phi_C)(\Phi_V(\mathbf{r}) - \phi_V)\right]$$

$$+ \frac{1}{V}\int_V d\mathbf{r}\left[-W_A(\mathbf{r})\Phi_A(\mathbf{r}) - W_C(\mathbf{r})\Phi_C(\mathbf{r}) - W_V(\mathbf{r})\Phi_V(\mathbf{r})\right]$$

$$- \frac{1}{V}\int_V d\mathbf{r}\left[\Xi(\mathbf{r})(1 - \Phi_A(\mathbf{r}) - \Phi_C(\mathbf{r}) - \Phi_V(\mathbf{r}))\right]$$

(7)

Here, $\varphi_{PLLA}$ and $\varphi_V$ are the overall volume fractions of PLLA and the voids, $\varphi_{PLLA} + \varphi_V = 1$; $\phi_A$, $\phi_C$, and $\phi_V$ are the overall average volume fractions of the A, C, and V species; $v_0$ is the volume of one lattice site, $V$ is the total volume of the system, $\chi_{ij}$ are the Flory-Huggins interaction parameters, $\Phi_i(\mathbf{r})$ is the local volume fraction of the i-th species, $W_i(\mathbf{r})$ is the conjugate "chemical potential" field, and $\Xi(\mathbf{r})$ is the Lagrange multiplier enforcing the "incompressibility" condition. The partition functions for PLLA and the voids, $Q_{PLLA}$ and $Q_V$, will be described below. For more detail on SCFT, see Fredrickson,[94] Matsen et al.,[91,92] and others.[95,96]

The partition functions for the PLLA and the voids are given by,

$$Q_{PLLA} = \int_V q^\dagger_{PLLA}(\mathbf{r},1) q_{PLLA}(\mathbf{r},1) d\mathbf{r} \qquad (8a)$$

$$Q_V = \int_V q^\dagger_V(\mathbf{r},1) q_V(\mathbf{r},1) d\mathbf{r} \qquad (8b)$$

The polymer (PLLA) and void (V) partition functions account for both conformational and translational entropy, as well as the enthalpic contributions. This is done by introducing the propagators $q_\alpha(\mathbf{r})$ and $q_\alpha^\dagger(\mathbf{r})$ ($\alpha$ = PLLA or V), determined based on the following partial differential equations,

$$\frac{\partial}{\partial s} q_\alpha(\mathbf{r},s) = \left[\nabla^2 - W_\alpha(\mathbf{r},s)\right] q_\alpha(\mathbf{r},s) \tag{9a}$$

$$\frac{\partial}{\partial s} q_\alpha^\dagger(\mathbf{r},s) = \left[\nabla^2 - W_\alpha(\mathbf{r},s)\right] q_\alpha^\dagger(\mathbf{r},s) \tag{9b}$$

with the boundary conditions, $q_\alpha(\mathbf{r},0)=1$ and $q_\alpha^\dagger(\mathbf{r},1)=1$. Here, the index $s$ denotes the position of a given repeat unit within a chain, with $s=0$ and $s=1$ corresponding to the two ends. The field $W_{PLLA}(\mathbf{r},s)$ equals to $W_A(\mathbf{r})$ for $0.25 < s < 0.75$ (amorphous block) and $W_C(\mathbf{r})$ otherwise (crystalline block).

The free energy (eq 7) must be minimized with respect to $\Phi_i(\mathbf{r})$, $W_i(\mathbf{r})$, and $\Xi(\mathbf{r})$ to obtain self-consistency equations which are then solved iteratively until convergence. Those self-consistency equations are,

$$\Phi_A(\mathbf{r}) = \frac{V}{Q_{PLA}} \int_V d\mathbf{r} \int_A ds \left[q^\dagger(\mathbf{r},s) q(\mathbf{r},s)\right] \tag{10a}$$

$$\Phi_C(\mathbf{r}) = \frac{V}{Q_{PLA}} \int_V d\mathbf{r} \int_C ds \left[q^\dagger(\mathbf{r},s) q(\mathbf{r},s)\right] \tag{10b}$$

$$\Phi_V(\mathbf{r}) = \frac{V}{Q_V} \exp\left[-W_V(\mathbf{r})\right] \tag{10c}$$

$$W_A(\mathbf{r}) = \chi_{AC} N\left(\Phi_C(\mathbf{r}) - \phi_C\right) + \chi_{AV} N\left(\Phi_V(\mathbf{r}) - \phi_V\right) + \Xi(\mathbf{r}) \tag{10d}$$

$$W_C(\mathbf{r}) = \chi_{AC} N\left(\Phi_A(\mathbf{r}) - \phi_A\right) + \chi_{CV} N\left(\Phi_V(\mathbf{r}) - \phi_V\right) + \Xi(\mathbf{r}) \tag{10e}$$

$$W_V(\mathbf{r}) = \chi_{AV} N\left(\Phi_A(\mathbf{r}) - \phi_A\right) + \chi_{CV} N\left(\Phi_C(\mathbf{r}) - \phi_C\right) + \Xi(\mathbf{r}) \tag{10f}$$

$$1 - \Phi_A(\mathbf{r}) - \Phi_C(\mathbf{r}) - \Phi_V(\mathbf{r}) = 0 \tag{10g}$$

In eq 10a, integration over s is in the domain *0.25 < s < 0.75* (amorphous block), while in eq 10b, the integration is over two crystalline blocks, *0 < s ≤ 0.25* and *0.75 ≤ s < 1*. Also note that all lengths are rescaled in units of $R_g = a\left(N/6\right)^{1/2}$, where *a* is the Kuhn length (or monomer size) and *N* is the number of Kuhn segments per PLLA chain.

To solve eqs (10a) – (10g), we use the following algorithm. At the beginning, the fields $W_i(\mathbf{r})$ are initialized using random Gaussian noise with small amplitude of 0.02. The "old" densities $\Phi_i^{old}(\mathbf{r})$ are set to equal to the average values of the species volume fractions, and the Lagrange multiplier $\Xi(\mathbf{r})=0$. The calculated densities, $\Phi_i^{calc}(\mathbf{r})$, are evaluated using eqs 8a – 8c. Then, the "new" densities are estimated as $\Phi_i^{new}(\mathbf{r}) = \lambda \Phi_i^{calc}(\mathbf{r}) + (1-\lambda)\Phi_i^{old}(\mathbf{r})$, where λ is a small number (here, we set λ = 0.04). (Note that the total amount of material of types A, V, and C is conserved – although the dynamics is formally non-conserved, the conservation stems from the fact that at each step, $\langle W_i \rangle = 0$.) Next, we update the pressure, using the formula, $\Xi^{new}(\mathbf{r}) = \Xi^{old}(\mathbf{r}) + \lambda \varepsilon \left[1 - \Phi_A(\mathbf{r}) - \Phi_C(\mathbf{r}) - \Phi_V(\mathbf{r})\right]$. Finally, we update the fields $W_i(\mathbf{r})$ using eqs 8d—8f, and repeat the process. The calculations stop after a prescribed convergence criterion is satisfied. Here, we chose $\langle |1 - \Phi_A - \Phi_C - \Phi_V| \rangle \leq 10^{-6}$.

We now describe our results in detail.

## 3. Results and Discussion

### 3.1. Bulk Relaxation Time and Free Volume

As the first step, we fit the measured *α*-relaxation time of amorphous PLLA to the VFTH form (eq 1). The data (blue circles) and the fit (orange line) are shown in Figure 2, and the fit parameters are summarized in Table 1.

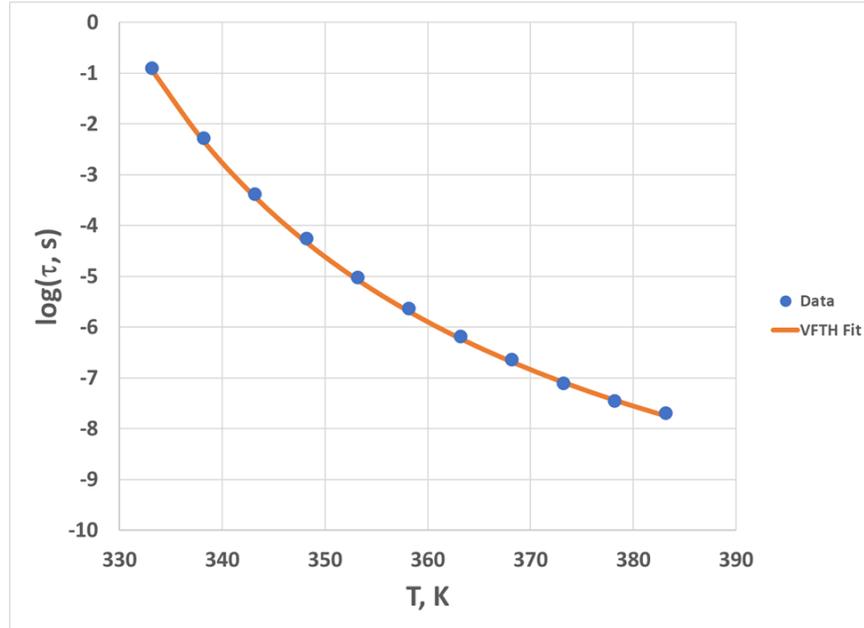

*Figure 2. Dielectric α-relaxation time as a function of temperature for amorphous PLLA. Blue circles are data from Cheng et al.[62] Orange line is the VFTH fit using parameters of Table 1.*

*Table 1. VFTH parameters for amorphous PLLA*

| Parameter | Value | Units |
|---|---|---|
| ln($\tau_\infty$,s) | -29.8 | |
| B | 1055 | K |
| $T_0$ | 295 | K |

Assuming B = 1055 K (Table 1) and α = 6.8x10$^{-4}$ K$^{-1}$ (the equilibrium rubbery-phase volumetric coefficient of thermal expansion of amorphous PLLA[97]), we obtain *A = 0.717*. The free volume as a function of temperature can be calculated according to eq 3 and is plotted in Figure 3. Interestingly, the fractional free volume at $T_g$ (defined as the point where $\log(\tau_\alpha) = 2$ and estimated to be close to 325 K) is 2.0%, close to the value estimated by Williams, Landel, and Ferry[9] (approximately 2.2%) and Lipson and White[18] (about 3%).

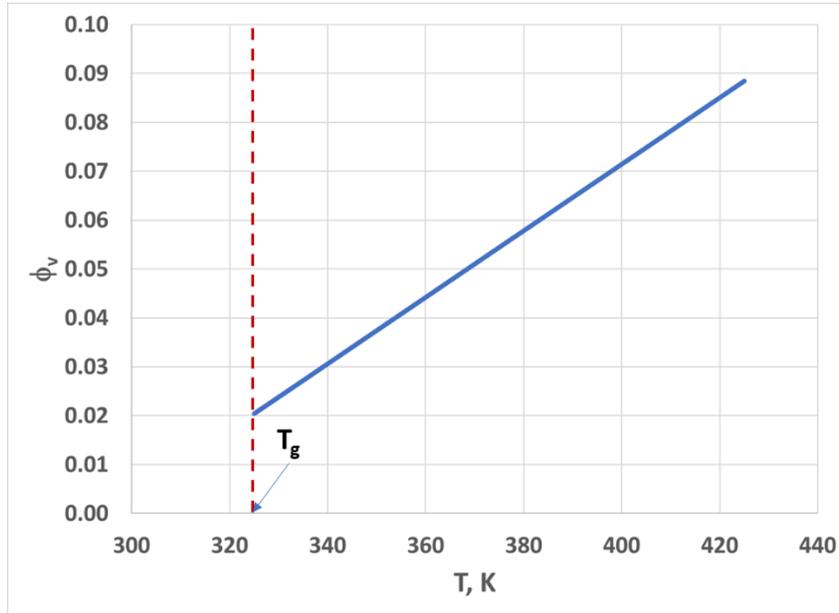

*Figure 3. Calculated fractional free volume of amorphous PLLA as function of temperature within the "naïve" FVT.*

### 3.2. Free Volume and Relaxation Time Profiles from Self-Consistent Field Theory

The structure of the semicrystalline PLLA, as modeled using SCFT, is pictured in Figure 4 (the profile is periodic so the location of the origin relative to the interface is arbitrary). The initial selection of the Kuhn length, $a$, and the incompatibility parameter, $\chi N$, was based on the requirement that the width of the amorphous domain, as defined below, equals to 10.6 nm; we thus used N = 200, $\chi$ = 0.339, and $a$ = 1.2 nm. The triblock is in the strong segregation region ($\chi N$ = 67.8), so the interfacial region ("interphase") is narrow. We chose to define the interphase boundaries as the points where $\min(\Phi_C, \Phi_A) = 1/e$, where $e$ is the base of natural logarithms. The distance between the two closest interphases is labeled $L$ and is identified with the width of the amorphous domain. In the following, all lengths will be scaled with respect to L.

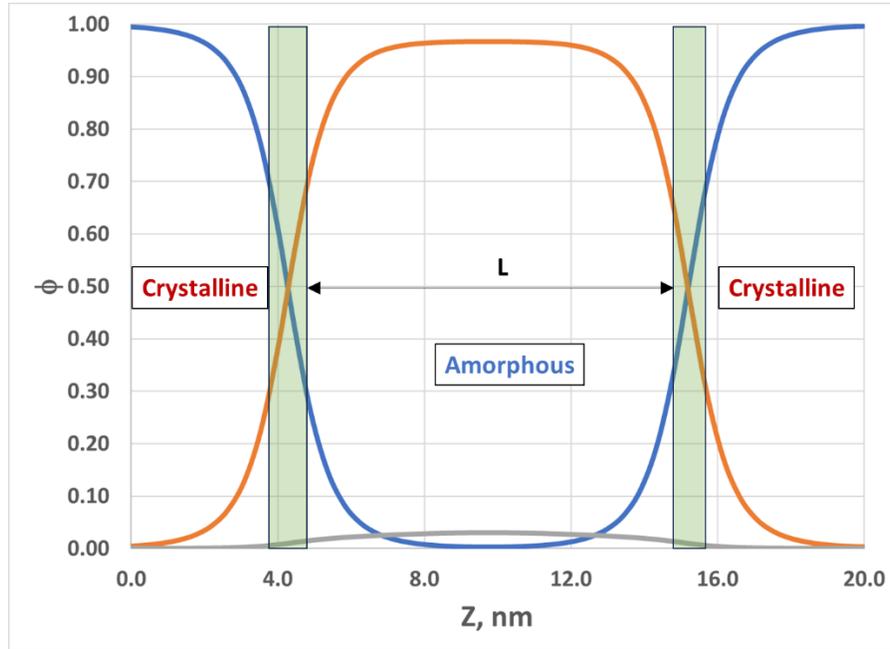

Figure 4. SCFT-calculated volume fractions of the amorphous PLLA (orange), crystalline PLLA (blue), and voids (grey) for T = 363 K. The green regions represent the "interphase", and L is the thickness of the amorphous domain.

To simplify the modeling, we assumed that the interaction parameter between the voids and the amorphous block, $\chi_{AV}$, is equal to zero, and the interaction parameter between the voids and the crystalline block, $\chi_{CV} = \chi_{AC}$. This corresponds to the scenario where the free volume is near-zero in the crystalline phase and is present only within the amorphous one – we believed it to be a reasonable assumption. Furthermore, we assumed that the Flory-Huggins parameter does not depend on temperature, at least in the narrow temperature interval studied here. Thus, in attempting to fit SCFT density profiles to the cHN free volume profiles, we varied only two parameters: $M$ (the "string" length) and $\varphi_V$ (the void volume fraction). The best model fits are given in Table 2. Note that the values of $\phi_V$ here are not the same as the bulk amorphous void fraction, $\phi_{V,b}$, given by eq 5. Rather, $\phi_V$ is the average volume fraction of the voids in the combined amorphous, interfacial, and crystalline domains. Thus, $\phi_V \leq 0.5\phi_{V,b}$. (To be more precise, we find $\phi_V \approx 0.44\phi_{V,b}$).

*Table 2. SCFT model parameters*

| T, K | $\phi_v$ | M |
|---|---|---|
| 348 | 0.0145 (± 0.0005) | 20 (±1) |
| 353 | 0.0165 (± 0.0005) | 17 (±1) |
| 363 | 0.0219 (± 0.0005) | 12 (±1) |

The results of the best-fit SCFT simulations are shown in Figure 5a, where the density profiles of the void component, $\Phi_v(z)$, are compared with the free volume profiles calculated using eq 6b. Alternatively, we can apply eq 6a to calculate the relaxation time based on the SCFT results and compare those to the estimates of Cheng and co-workers[62] (Figure 5b). We can see that the two analyses result in very similar behaviors.

The temperature dependence of the "string length", *M*, is related to the broadening of the correlation length, $\xi$, upon lowering the temperature. Douglas and co-workers[87] showed that the string length *M* is proportional to the cooperatively rearranging volume (CRR); they also demonstrated that *M* decreases upon increasing temperature and that behavior can be approximately captured by the Arrhenius law. Our modeling agrees with that analysis, with *log(M)* scaling linearly with *1/T*, although the temperature range is relatively small so other functional forms could also apply.

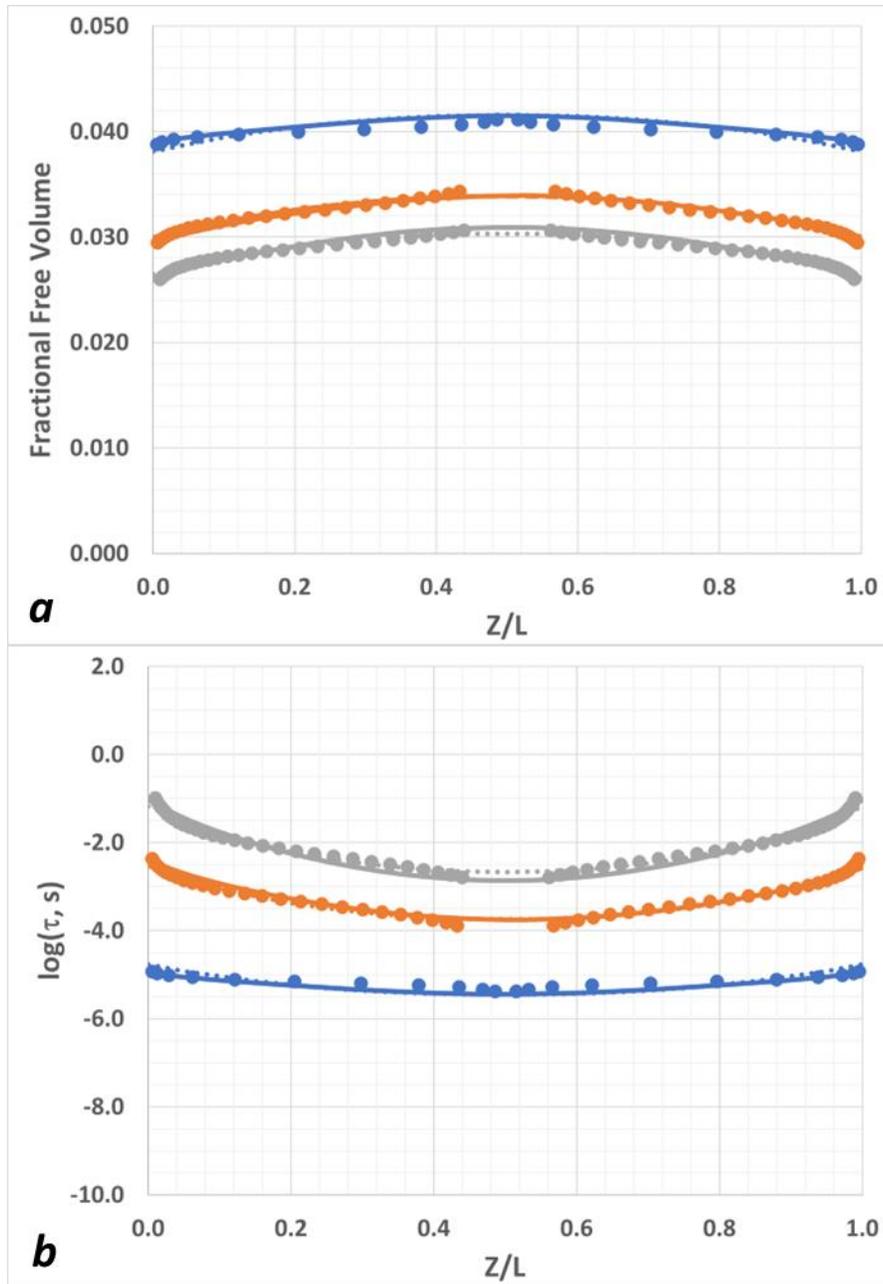

*Figure 5. (a) Estimated free-volume profiles from SCFT (solid lines), the phenomenological model (dashed lines), and based on the dielectric measurement data analysis (circles). (b) Logarithm of α-relaxation time, from SCFT (solid lines), the phenomenological model (dashed lines), and from dielectric measurements (circles). The SCFT and phenomenological models are practically indistinguishable, except for small differences near the surfaces for T = 363 K and in the middle of the amorphous domain for T = 348 K. Grey lines and circles correspond to T = 348 K, orange to T = 353 K, and blue to T = 363 K.*

### 3.3. Phenomenological Modeling of the Relaxation Time and Free-Volume Profiles for Semicrystalline PLLA

The SCFT approach described the density profiles and the relaxation times based on a simple coarse-grained model; however, this is a numerical solution of nonlinear partial differential equations which is often difficult to implement or reproduce. We now propose a simple phenomenological analytical description that can be compared with the SCFT solution or the cHN analysis. Specifically, we assume that,

$$\phi_v(z;T) = \phi_{v,b}(T)\left[1 + q\left(\exp\left[-\frac{z}{\xi}\right] + \exp\left[-\frac{L-z}{\xi}\right]\right)\right] \quad (11)$$

Here, $\phi_{v,b}(T)$ is the bulk fractional free volume of the amorphous PLLA, given by eq 5; $\xi$ is the correlation length, and $q$ is the free volume difference between the pure amorphous material and the same material near the crystal surface. Assuming that the correlation length is temperature-independent, we fit the experimental free-volume profiles for the three temperatures. Those profiles are generated by taking the cHN-calculated local relaxation times (see Figure 7b of Cheng et al.[62]) and applying eq 6b to them. The fitting is done using Excel Generalized Reduced Gradient (GRG) minimization. The results are shown in Figure 5 (dashed lines) and the model parameters in Table 3.

Table 3. Model parameters used to fit the FFV and relaxation time profiles (see eq 9).

| T, K | $\phi_{v,b}$ | $\xi$, nm | q |
|---|---|---|---|
| 348 | 0.036 | 4.8 (±0.2) | -0.24 (±0.010) |
| 353 | 0.039 | 4.7 (±0.2) | -0.22 (±0.008) |
| 363 | 0.046 | 4.6 (±0.2) | -0.16 (±0.007) |

We can now extrapolate the model parameters to other temperatures. Since the dependence of $\phi_{v,b}(T)$ is already known (see eq 3), and $\xi$ is approximately temperature-independent, we need only to calculate the T-dependence of $q$. It is evident that as T → $T_m$, q → 0 (since after the melting, there are no more crystalline domains and crystal-amorphous interfaces). We then assume that $q$ is a function of ($T_m/T$ – 1) that can be approximated by a power-law,

$$q = -K \left[ \frac{T_m}{T} - 1 \right]^m \qquad (12)$$

Here, $K$ = 2.385 (±0.005), and $m$ = 1.78 (±0.01); $T_m$ = 436 K based on the differential scanning calorimetry (DSC) measurements.[62] We can now calculate the average relaxation time for other temperatures in the $T_g < T < T_m$ range and compare with experimental data (Figure 6). The agreement is very good, especially given the simplicity of the underlying model. The relaxation time of a semi-crystalline PLLA can be orders of magnitude larger than that of its amorphous counterpart at the same temperature. The difference becomes smaller and ultimately vanishes as the material is heated towards its melting point.

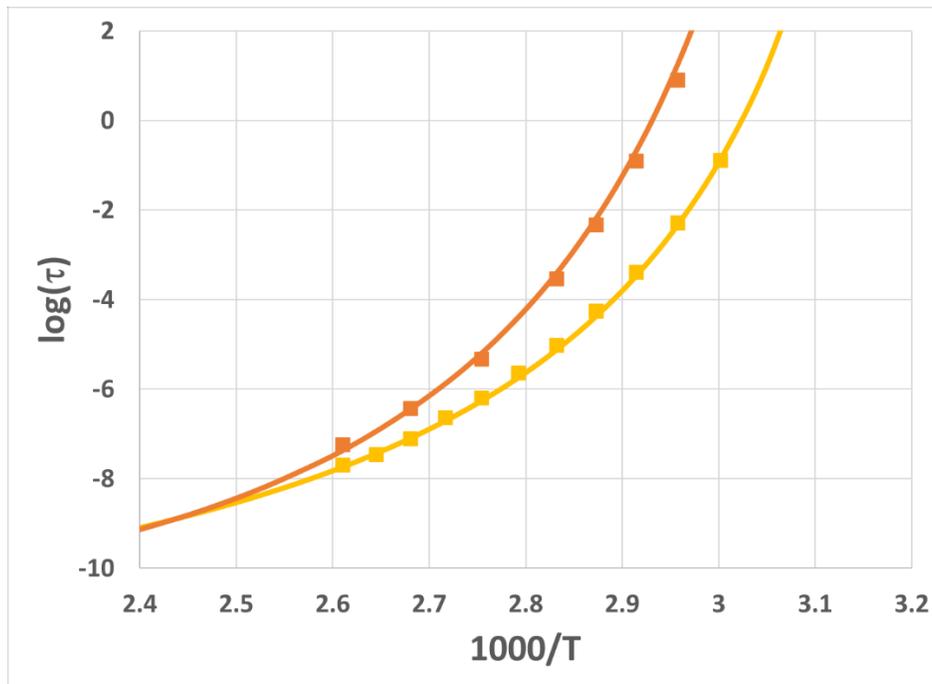

Figure 6. Experimental (squares) and model-fit (lines) relaxation time as function of temperature. Gold – pure amorphous PLLA, orange – semi-crystalline PLLA. Data are from Cheng et al.[62]

The above analysis suggests that the correlation length in this system is about 4.8 nm or roughly four Kuhn lengths ($l_K \approx 1.1 – 1.2$ nm). This number is consistent with other estimates of polymer correlation lengths (typically in the range 3-5 nm).[62,98–100]

### 3.4. Discussion

The proposed approach (SCFT combined with "naïve" FVT) relates the structure and the dynamics of semicrystalline polymers between their glass transition and melting temperatures. The free volume distribution is modeled using a lattice model, based on the ideas of Sanchez and Lacombe[75,76,84] and Hong and Noolandi.[68] The free volume holes are strongly repelled from the crystalline domains, due to the higher cohesive energy density of the crystalline, compared to the amorphous, polymer. The density profile of the holes is a monotonically decreasing function of the distance from the amorphous-crystalline surface. The maximum local free volume content in the center of the amorphous region is significantly larger than near the interfaces, resulting in the significant slowing down of all the motion at the interfaces (by about 1-2 orders of magnitude), which is similar to what is observed in supported thin films or nanocomposites.[64,101–103]

The proposed analysis uses three simplifying assumptions. The first is that the density profile determining the dielectric relaxation is fully equilibrated. This assumption would obviously break down as one approaches $T_g$ from above, but should be reasonably accurate at temperatures above, say $T_g + 5K$. The second assumption is that all the changes in the specific volume are associated with the free volume, so the rate of the free volume increase with temperature is given by the experimentally observed coefficient of thermal expansion. This assumption is likely incorrect, in a sense that the specific volume change with temperature should have both the "free volume" and "occupied volume" contributions (see, e.g., Lipson and White[18,57,82] or Ginzburg et al.[13–15]) However, within a narrow temperature range, one could assume the two parts of the volume change to be linearly proportional; thus, while eq 3 still holds, one will need to write, $A/B = \alpha k$, where *k* is the fraction of the relative specific volume change due to the free volume change. The last assumption is, of course, the equation 4 itself (the Doolittle or "naïve" free-volume theory). In effect, it assumes that the "local" relaxation time is a function of the "local" structure and not some strongly cooperative effects (see, e.g., Schweizer and co-workers,[30,32–34,53] Lipson and co-workers,[57,59,80–82] and others). In a way, the question is whether the coarse-grained "local free volume" $\phi_v(z)$ is sufficient to determine the local relaxation time $\tau_\alpha(z)$ – or one needs to also include the contributions from the "neighboring layers". Here, we attempted to address the challenge of incorporating the correlations by modeling the free volume voids as "homopolymers", rather than "solvents", within the SCFT framework. This approach conforms with the ideas of Douglas et al.[87–90] according to which the free volume holes (regions of local high mobility) associate into "springs" with temperature-dependent length. Thus, the free energy functional includes long-range correlations due to the void "connectivity". An alternative description of the same long-range correlation could include instead the addition of higher-order spatial derivatives in the density functional.[63] Overall, developing a molecular-informed, accurate field-theoretical approach to describe the free volume profiles in polymer thin films, nanocomposites, and semi-crystalline polymers is still a major challenge that needs to be addressed.

## 4. Conclusions

We combined two models, the free-volume theory and the compressible Self-Consistent Field Theory, to describe the relaxation time profiles within the amorphous-crystalline "interphase" in semi-crystalline polymers at temperatures between the glass transition and melting temperatures. The combined approach can successfully explain the relaxation time distribution profile obtained through dielectric spectra via a continuous Havriliak-Negami analysis and leads to a description of the free-volume profiles in the amorphous phase of a semicrystalline PLLA. We believe the model can be used to describe the structure and dynamics of many other inhomogeneous systems above $T_g$, from thin films (both free-standing and supported) to other semi-crystalline materials to segmented block copolymers to nanocomposites. This will be the subject of future work.

## Acknowledgments

The author thanks Shiwang Cheng (MSU) for providing the dielectric spectroscopy data, for many fruitful discussions and helpful suggestions, and for providing a detailed description of the continuous Havriliak-Negami analysis.